\documentstyle[12pt]{article}

\newcommand{\EQ}{\begin{equation}}
\newcommand{\EN}{\end{equation}}
\newcommand{\bea}{\begin{eqnarray}}
\newcommand{\ena}{\end{eqnarray}}
\newcommand{\vs}[1]{\vspace{#1 mm}}

\renewcommand{\b}{\beta}
\renewcommand{\c}{\gamma}
\renewcommand{\d}{\delta}
\newcommand{\e}{\epsilon}
\renewcommand{\t}{\theta}
\newcommand{\tb}{{\bar \theta}}
\newcommand{\shalf}{\frac{1}{2}}
\newcommand{\pa}{\partial}

\newcommand{\dZ}{\frac{dzd^2\t}{2\pi i}}
\newcommand{\nn}{\nonumber \\}

\begin{document}

\topmargin 0pt
\oddsidemargin 5mm

\renewcommand{\Im}{{\rm Im}\,}
\newcommand{\NP}[1]{Nucl.\ Phys.\ {\bf #1}}
\newcommand{\PL}[1]{Phys.\ Lett.\ {\bf #1}}
\newcommand{\CMP}[1]{Comm.\ Math.\ Phys.\ {\bf #1}}
\newcommand{\PR}[1]{Phys.\ Rev.\ {\bf #1}}
\newcommand{\PRL}[1]{Phys.\ Rev.\ Lett.\ {\bf #1}}
\newcommand{\PTP}[1]{Prog.\ Theor.\ Phys.\ {\bf #1}}
\newcommand{\PTPS}[1]{Prog.\ Theor.\ Phys.\ Suppl.\ {\bf #1}}
\newcommand{\MPL}[1]{Mod.\ Phys.\ Lett.\ {\bf #1}}
\newcommand{\IJMP}[1]{Int.\ Jour.\ Mod.\ Phys.\ {\bf #1}}

\begin{titlepage}
\setcounter{page}{0}
\begin{flushright}
OU-HET 210 \\
hep-th/9504099
\end{flushright}

\vs{8}
\begin{center}
{\Large UNIVERSAL STRING AND SMALL $N=4$ SUPERSTRING}

\vs{20}
{\large Nobuyoshi Ohta\footnote{e-mail address:
ohta@phys.wani.osaka-u.ac.jp}
and Takashi Shimizu\footnote{e-mail address:
simtak@phys.wani.osaka-u.ac.jp}}\\
{\em Department of Physics, Osaka University, Toyonaka, Osaka 560, Japan}
\end{center}

\vs{30}
\centerline{{\bf{Abstract}}}

It was previously shown that most of the superstrings can be obtained
from those with higher world-sheet supersymmetry as spontaneously
broken phases. In this  paper, we show that the small $N=4$ superstring,
which was left out of this hierarchy of the universal string, can
be obtained from the large $N=4$ strings. We also show that the $N=2$
string is a special vacuum of small $N=4$ string. Thus all the known
superstring theories can be derived from a universal string by
spontaneous breakdown of supersymmetry.

\end{titlepage}
\newpage
\renewcommand{\thefootnote}{\arabic{footnote}}
\setcounter{footnote}{0}

Recently it has been shown that any $N=0~(N=1)$ string can be obtained
from $N=1~(N=2)$ superstring as a spontaneously broken phase~\cite{BV}.
This remarkable discovery has opened the possibility of constructing
a universal string from which all known string theories can be
derived just by selecting different vacua. Indeed this has been
further generalized, and it is now known that an arbitrary $N$-extended
superstring can be regarded as a special class of vacua for $(N+1)$
superstring~\cite{BOP}.

At the moment, this hierarchy has been successfully formulated
for $N$-extended superstrings based on linear superconformal algebras
(SCAs) with $SO(N)$ current algebra~\cite{ADE}. However there is an
important class of superstrings which is usually termed the `small'
$N=4$ superstring with only $SU(2)$ current algebra. This class of
superstrings is the largest extended
one that has been formulated by lagrangian and also constitutes an
interesting theory from the point of view of topological string.
It is thus important to examine if this class of superstrings also
belongs to the hierarchy of the universal string or if it is completely
decoupled from such a unified approach.

The purpose of this paper is to show that it is in fact possible to
embed the small $N=4$ superstring into the hierarchy. We will show
that the $N=2$ superstring is realized as a special vacuum of the
small $N=4$ superstring, which in turn is realized in the large
$N=4$ superstring. We also point out that the large $N=4$ superstring
with an arbitrary parameter $x$ (the ratio of the levels of $SU(2)$
current algebras contained in the theory) is realized in the already
existing embeddings. This completes the program of the universal
string concerning the linear SCA~\cite{BOP}.

In order to discuss the embeddings of some superstring theory into
the small $N=4$, we must first ask what theory can be embedded into
the small $N=4$ string. This can be easily answered if one notices that
the $N=3$ superstring contains a symmetry generator of dimension
$\shalf$ which is not present in the small $N=4$ superstring.
Consequently the largest super-subalgebra of the small $N=4$ SCA is
the $N=2$ SCA. We thus conclude that it is only possible to embed
the $N=2$ superstring into the small $N=4$ string.

Let us first discuss the embedding of $N=2$ superstrings into the small
$N=4$ one. For this purpose, we use $N=2$ superfields. Thus the small
$N=4$ SCA can be written in the $N=2$ superfields as follows:
\bea
T(Z_1) T(Z_2) &\sim& \frac{\frac{1}{3}c + \t_{12}\tb_{12} T}{z_{12}^2}
  +\frac{-\t_{12}DT + \tb_{12}{\bar D}T + \t_{12}\tb_{12}\pa T}{z_{12}}, \nn
T(Z_1)G_c(Z_2) &\sim&  \left( \frac{ \t_{12}\tb_{12} }{z_{12}^2}
+ \frac{2}{z_{12}} \right)  G_c
  +\frac{ \tb_{12}{\bar D}G_c + \t_{12}\tb_{12}\pa G_c}{z_{12}}\ , \nn
T(Z_1)G_a(Z_2) &\sim&  \left( \frac{ \t_{12}\tb_{12} }{z_{12}^2}
- \frac{2}{z_{12}} \right)  G_a
  +\frac{- \t_{12} DG_a + \t_{12}\tb_{12}\pa G_a}{z_{12}}\ , \nn
G_c(Z_1) G_a(Z_2) &\sim& - \frac{c}{6}
\left( \frac{ \t_{12}\tb_{12} }{z_{12}^3} - \frac{1}{z_{12}^2} \right)
- \shalf
\left( \frac{ \t_{12}\tb_{12} }{z_{12}^2} - \frac{2}{z_{12}} \right)
T +\frac{ \tb_{12}}{z_{12}} {\bar D}T \ ,
\label{sn4}
\ena
where $G_c$ and $G_a$ are chiral and antichiral superfields,
respectively. Here $Z \equiv (z,\t,\tb)$, and $D \equiv \pa_\t -\tb
\shalf \pa_z, {\bar D} \equiv \pa_\tb - \t \shalf \pa_z$ are
the covariant derivatives, and $z_{12} \equiv z_1 - z_2 + \shalf(\t_1
\tb_2+\tb_1\t_2), \t_{12} \equiv \t_1-\tb_2, \tb_{12}\equiv
\tb_1-\tb_2$. The $N=2$ subalgebra is given by the first line
of eq.~(\ref{sn4}) with the generator $T$ alone.

In order to embed the $N=2$ superstring into the small $N=4$, we
take an arbitrary background $T_m$ for the $N=2$ superstring which
satisfies the first relation in eq.~(\ref{sn4}). We then introduce two
sets of chiral and antichiral fields $(\xi_c,\b_c), (\xi_a,\b_a)$,
corresponding to the generators $G_c$ and $G_a$.
Here $\xi_c,\xi_a$ are anticommuting fields of spin $-\shalf$ while
$\b_c,\b_a$ are commuting ones of spin $1$ with the operator product
expansion (OPE)
\EQ
\xi_c(Z_1) \b_c(Z_2) \sim \frac{\tb_{12}}{z_{12}}, \qquad
\xi_a(Z_1) \b_a(Z_2) \sim \frac{\t_{12}}{z_{12}}.
\EN
These fields are regarded as part of matter fields of our theory.
We then find the following (non-critical) realization satisfies
the small $N=4$ SCA (\ref{sn4}) with the central charge shifted to
$c-18$:
\bea
T &=& T_m + 2 {\bar D}\xi_c \b_c - \xi_c {\bar D}\b_c
 - 2 D\xi_a \b_a + \xi_a \b_a, \nn
G_c &=& \b_c + D \left[ \xi_a T_m - \xi_a\xi_c{\bar D}\b_c + 2 \xi_a
{\bar D}\xi_c \b_c - \xi_a D\xi_a \b_a + \frac{18-c}{6}
\pa\xi_a \right] ,\nn
G_a &=& \b_a.
\label{sn4gen}
\ena
The critical algebra is for $c=6$. However the non-critical
realization may be useful in analyzing the properties of topological
strings.

It should be noted that the structure of the additional
super-generator $G_c$ is completely different from the known
embeddings~\cite{BV,BOP}. In these cases, it was obtained just by
``improving'' the BRST current by total derivatives such that it
satisfies the correct OPE. However our generator $G_c$ has a
fundamentally different structure from the BRST current.

Substituting the generators (\ref{sn4gen}) for $c=6$ into the BRST
operator for the small $N=4$ string
\bea
Q_{sN4} &=& \oint \dZ \left[ C_t \left( T + \shalf \pa C_t B_t
 + \shalf D C_t {\bar D} B_t + \shalf {\bar D} C_t D B_t
 + 2 D C_a B_a + C_a DB_a \right.\right.\nn
&& \left.\left. - 2 {\bar D} C_c B_c
 - C_c {\bar D} B_c \right) -B_t C_c C_a \right]
 + \oint \frac{dz d\tb}{2\pi i} C_c G_c
 + \oint \frac{dz d\t}{2\pi i} C_a G_a,
\ena
we find
\bea
Q_{sN4} &=& \oint \dZ \left[ C_t \left( T_m + \shalf \pa C_t B_t
 + \shalf D C_t {\bar D} B_t + \shalf {\bar D} C_t D B_t
 + 2 D C_a B_a + C_a DB_a \right. \right. \nn
&& \left. - 2 {\bar D} C_c B_c - C_c {\bar D} B_c
 - 2 D \xi_a \b_a + \xi_a D\b_a + 2 {\bar D} \xi_c \b_c
 - \xi_c {\bar D} \b_c \right) - B_t C_c C_a \nn
&& \left. + C_c \xi_a T_m - C_c \xi_a\xi_c{\bar D}\b_c
 + 2 C_c \xi_a{\bar D} \xi_c \b_c
 - C_c \xi_a D \xi_a \b_a + 2 C_c \pa \xi_a  \right] \nn
&& + \oint \frac{dz d\tb}{2\pi i} C_c \b_c
 + \oint \frac{dz d\t}{2\pi i} C_a \b_a.
\label{sn4brs}
\ena

We can show the equivalence of this particular class of the small
$N=4$ string to the $N=2$ string. Our method to show this is to use
a similarity transformation that maps this BRST operator (\ref{sn4brs})
for the small $N=4$ superstring to a sum of those for the $N=2$ and
topological sectors~\cite{R}. We find that the following similarity
transformation does this job:
\EQ
e^{R_3}e^{R_2}e^{R_1}Q_{sN4}e^{-R_1}e^{-R_2}e^{-R_3}
= Q_{N=2} + Q_{top},
\label{n2top}
\EN
where
\bea
Q_{N=2} &=& \oint \dZ  C_t \left( T_m + \shalf \pa C_t B_t
 + \shalf D C_t {\bar D} B_t + \shalf {\bar D} C_t D B_t \right), \nn
Q_{top} &=& \oint \frac{dz d\tb}{2\pi i} C_c \b_c
 + \oint \frac{dz d\t}{2\pi i} C_a \b_a,
\ena
and where
\bea
R_1 &=& - \oint \dZ C_c \xi_a B_t, \nn
R_2 &=& - \oint \dZ C_c \xi_a D \xi_a B_a, \nn
R_3 &=& \oint \dZ [ D C_t \xi_a B_a + C_t D \xi_a B_a
 - {\bar D} C_t \xi_c B_c - C_t {\bar D} \xi_c B_c ].
\ena

With this form (\ref{n2top}) of the BRST operator, it is obvious that
the cohomology of the $Q_{sN4}$ is a direct product of those of
$Q_{N=2}$ and $Q_{top}$. The BRST operator $Q_{top}$ imposes the
condition that the fields $(C_c,B_c,\xi_c,\b_c)$ as well as
$(C_a,B_a,\xi_a,\b_a)$ decouple from the physical sector and the
cohomology of the operator consists only of their vacuum. Thus
we obtain one-to-one correspondence of the cohomologies of
$Q_{sN4}$ and $Q_{N=2}$, and the small $N=4$ string propagating in
the background described by (\ref{sn4gen}) is equivalent to the
$N=2$ string. It has also been shown that this phenomenon can be
interpreted as a spontaneous breakdown of superconformal
symmetry~\cite{K}.

Let us next turn to the embedding of the small $N=4$ into the large
$N=4$ superstring. In order to discuss this embedding, it turns out
convenient to express the $N=4$ SCA in $N=1$ superfields~\cite{n1s}:
\bea
T(Z_1){\cal O}(Z_2) &\sim& \frac{\frac{c}{6}}{z_{12}^3} T
  +\frac{3}{2} \frac{\t_{12}}{z_{12}^2} T + \shalf \frac{1}{z_{12}}
 DT + \frac{\t_{12}}{z_{12}}\pa T, \nn
T(Z_1){\cal O}(Z_2) &\sim& h_{\cal O} \frac{\t_{12}}{z_{12}^2} {\cal O}
 + \shalf \frac{1}{z_{12}} D{\cal O}
 + \frac{\t_{12}}{z_{12}}\pa{\cal O}, \nn
T(Z_1)H(Z_2) &\sim& \frac{\frac{c}{12}+\shalf \t_{12} H}{z_{12}^2}
 + \shalf \frac{1}{z_{12}} DH + \frac{\t_{12}}{z_{12}}\pa H, \nn
C_3(Z_1)C_\pm(Z_2) &\sim& \pm\frac{2 C_\pm+\t_{12}DC_\pm}{z_{12}},\qquad
C_3(Z_1)C_3(Z_2) \sim \frac{\frac{c}{3}}{z_{12}^2}
  +\frac{\t_{12} 2 T}{z_{12}},\nn
C_+(Z_1)C_-(Z_2) &\sim& \frac{\frac{2}{3}c}{z_{12}^2}
  +\frac{4C_3+\t_{12}(4T+2DC_3)}{z_{12}},\nn
N_3(Z_1) C_\pm(Z_2) &\sim& N_\pm(Z_1) C_3(Z_2) \sim
  \mp\frac{N_\pm+\t_{12}C_\pm}{z_{12}},\nn
N_3(Z_1) C_3(Z_2) &\sim& -\frac{\frac{c}{6}\t_{12}}{z_{12}^2}-
 \frac{H}{z_{12}},\qquad
N_\pm(Z_1) C_\mp(Z_2) \sim \frac{\frac{c}{3}\t_{12}}{z_{12}^2}+
 \frac{2(H\mp N_3)\pm\t_{12}2C_3}{z_{12}},\nn
N_3(Z_1) N_\pm(Z_2) &\sim& \frac{\mp \t_{12}}{z_{12}}N_\pm,\qquad
N_+(Z_1) N_-(Z_2) \sim \frac{k}{z_{12}}-\frac{\t_{12}}{z_{12}}2N_3,\nn
N_3(Z_1) N_3(Z_2) &\sim& \frac{\frac{k}{2}}{z_{12}},\qquad
H(Z_1) C_\pm(Z_2) \sim \frac{N_\pm}{z_{12}},\nn
H(Z_1) C_3(Z_2) &\sim& -\frac{N_3}{z_{12}},\qquad
H(Z_1) H(Z_2) \sim -\frac{\frac{k}{2}}{z_{12}},
\label{n1large}
\ena
where $Z \equiv (z,\t)$, $D \equiv \pa_\t +\t \pa_z$ is the $N=1$
covariant derivative, $z_{12} \equiv z_1 - z_2 + \t_1\t_2, \t_{12}\equiv
\t_1-\t_2$ and the operators $\cal O$ stand for $C_\pm, C_3, N_\pm,
N_3$ with their dimensions $h_C=1, h_N=\shalf$, respectively.\footnote{
The component structure of the $N=1$ superfields in the notation
of ref.~\cite{BO1} is as follows:
\begin{eqnarray*}
T &=& \shalf ( G_{(1,2)}+G_{(2,1)}) + \t T, ~~
C_+ = 2 H_+ + 2 \t [G_{(1,1)}+(1+x)F'_{(1,1)}], \\
C_- &=& 2 H_- - 2 \t [G_{(2,2)}+(1+x)F'_{(2,2)}], ~~
C_3 = 2 H_3 + \t [G_{(2,1)}-G_{(1,2)}-(1+x)F'_{(1,2)}
 + (1+x)F'_{(2,1)}], \\
N_+ &=& 2 F_{(1,1)} + \t (H_+ - J_+), ~~
N_- = -2 F_{(2,2)} + \t (H_- - J_-), \\
N_3 &=& F_{(1,2)} - F_{(2,1)} - \t (H_3 + J_3), ~~
H = F_{(1,2)} + F_{(2,1)} + \t J,
\end{eqnarray*}
where $G, H, J$ and $F$ are the super generators, two $SU(2)$
currents and fermionic currents, respectively.}
Here $T,N,H$ are fermionic superfields, $C$ are bosonic
and the central charge $c$ and $k$ are given as
\EQ
c=3 k_1, \qquad
k=k_1+k_2,
\EN
in terms of the levels $k_1,k_2$ of the two $SU(2)$ current algebras
in the large $N=4$ SCA. Notice that the generators $T,C$ make a closed
algebra which is the small $N=4$ SCA. It is interesting to notice
that the algebra can be expressed without explicit dependence on
the ratio of the levels of current algebras.

We note that here is a very special situation different from all
embeddings known so far~\cite{BV,BOP}. Since the small $N=4$ SCA is a
subalgebra of the large $N=4$, the OPEs of unbroken generators with
unbroken ones close as usual.\footnote{We are referring as
``unbroken'' to those generators that remain exact in the broken phase
($N=2$ superconformal generator $T$ in our previous example) and to
other generators as ``broken''.} In the usual embeddings, the OPEs of
unbroken generators with broken ones give broken ones. However, in our
special case, the OPE of $C$ with $N$ involves $C$. This makes
our embedding very special.

In the usual case, all the unbroken generators are again themselves
unbroken generators in the theory with higher symmetries. However,
here this cannot be the case because they are independent of the
additional matter fields and cannot have nonzero OPE with $N$
generators which consist of solely new additional matter fields.
It turns out that the generators $C$ need an infinite number
of terms dependent on the new fields, just as the non-critical
embeddings found in ref.~\cite{BO}. Remarkably we can determine the
explicit structure of these infinite terms from the consistency
of the OPE (\ref{n1large}).

Now our realization is as follows. We take generators $T_m,C_m$
which satisfy the small $N=4$ SCA in (\ref{n1large}) with $c=-12$ and
introduce additional matter fields $(b^+, \c_+), (b^-, \c_-),\\
(b^3, \c_3), (b^0,\c_0)$ of spins $(\shalf,0)$. They have the OPE
\EQ
\c_i(Z_1) b^j (Z_2) \sim \d_i^j \frac{\t_{12}}{z_{12}}.
\EN
Here and in what follows, we denote commuting fields by $\b,\c$ and
anticommuting ones by $b,c$ unless otherwise stated.
We then find that the following generators satisfy the large $N=4$
algebra (\ref{n1large}) with $c=0,k=0$:
\bea
T &=& T_m  + \shalf b^+ \pa\c_+ +  \shalf D b^+ D\c_+
 + \shalf b^- \pa\c_- + \shalf D b^- D\c_- \nn
&& + \shalf b^3 \pa\c_3 + \shalf D b^3 D\c_3
 + \shalf b^0 \pa\c_0 + \shalf D b^0 D\c_0
 + \sum_{n=1}^{\infty} T_n, \nn
C_+ &=& C_{m,+} + 2 b^- D\c_3 + 2 b^0 D\c_+
 + \sum_{n=1}^{\infty} C_{+,n} , \nn
C_- &=& C_{m,-} + 2 b^3 D\c_- + 2 b^+ D\c_0
 + \sum_{n=1}^{\infty} C_{-,n} , \nn
C_3 &=& C_{m,3} - b^+ D\c_+ + b^- D\c_- - b^3 D\c_3 + b^0 D\c_0
 + \sum_{n=1}^{\infty} C_{3,n} , \nn
N_+ &=& b^- - b^- \c_3 - b^0 \c_+ , \nn
N_- &=& b^+ - b^+ \c_0 - b^3 \c_- , \nn
N_3 &=& \shalf (b^3 - b^0 - b^+ \c_+ + b^- \c_-
 - b^3 \c_3 + b^0 \c_0) , \nn
H &=& \shalf (- b^3 - b^0 + b^+ \c_+ + b^- \c_-
 + b^3 \c_3 + b^0 \c_0) ,
\label{n1gen}
\ena
where
\bea
T_n &=& - D\pa \left[
 \frac{1}{n} \c_3^n + \frac{1}{n} \c_0^n
 + \sum_{l,k} \frac{l}{(l!)^2} \frac{(k+l-1)!(n-k-l-1)!}{k!(n-k-2l)!}
 (\c_+ \c_-)^l \c_3^{n-2l-k}\c_0^k \right] , \nn
C_{+,n} &=&
 \left[ \c_0^n - \c_0^{n-1}\c_3 + (n-1)\c_+ \c_- \c_0^{n-2} \right. \nn
&& \left. \qquad + \sum_{l,k} \frac{l(l-1)}{(l!)^2}
 \frac{(k+l-2)!(n-k-l)!}{k!(n-k-2l)!}
 (\c_+ \c_-)^l \c_0^{n-2l-k}\c_3^k \right] C_{m,+} \nn
&& - \sum_{l,k} \frac{1}{(l!)^2} \frac{(k+l)!(n-2-k-l)!}{k!(n-2-k-2l)!}
 \c_+^{l+2} \c_-^l \c_3^{n-2-2l-k}\c_0^k  C_{m,-} \nn
&& - 2\c_+ \left[ \c_0^{n-1}
 + \sum_{l,k} \frac{l}{(l!)^2} \frac{(k+l-1)!(n-1-k-l)!}{k!(n-1-k-2l)!}
 (\c_+ \c_-)^l \c_0^{n-1-2l-k}\c_3^k \right] C_{m,3} \nn
&& -4 (\pa \c_+) \c_0^{n-1}
 - \sum_{l,k} \frac{4}{(l!)^2} \frac{(k+l)!(n-2-k-l)!}{k!(n-2-k-2l)!}
 \c_+^{l+1} \c_-^l (\pa\c_3)\c_3^{n-2-2l-k}\c_0^k \nn
&& - \sum_{l,k} \frac{4l}{(l!)^2} \frac{(k+l)!(n-2-k-l)!}{k!(n-1-k-2l)!}
 (\c_+ \c_-)^l (\pa\c_+)\c_3^{n-1-2l-k}\c_0^k, \nn
C_{-,n} &=&
 - \sum_{l,k} \frac{1}{(l!)^2} \frac{(k+l)!(n-2-k-l)!}{k!(n-2-k-2l)!}
 \c_+^{l} \c_-^{l+2} \c_0^{n-2-2l-k}\c_3^k  C_{m,+} \nn
&& + \left[ \c_3^n - \c_3^{n-1}\c_0 +(n-1)\c_+\c_- \c_3^{n-2} \right. \nn
&& \left. \qquad + \sum_{l,k} \frac{l(l-1)}{(l!)^2}
 \frac{(k+l-2)!(n-k-l)!}{k!(n-k-2l)!}
 (\c_+ \c_-)^l \c_3^{n-2l-k}\c_0^k \right] C_{m,-} \nn
&& + 2\c_- \left[ \c_3^{n-1}
 + \sum_{l,k} \frac{l}{(l!)^2} \frac{(k+l)!(n-2-k-l)!}{k!(n-1-k-2l)!}
 (\c_+ \c_-)^l \c_0^{n-1-2l-k}\c_3^k \right] C_{m,3} \nn
&& -4 (\pa \c_-) \c_3^{n-1}
 - \sum_{l,k} \frac{4}{(l!)^2} \frac{(k+l)!(n-2-k-l)!}{k!(n-2-k-2l)!}
 \c_+^l \c_-^{l+1} (\pa\c_0)\c_0^{n-2-2l-k}\c_3^k \nn
&& - \sum_{l,k} \frac{4l}{(l!)^2} \frac{(k+l)!(n-2-k-l)!}{k!(n-1-k-2l)!}
 (\c_+ \c_-)^l (\pa\c_-)\c_0^{n-1-2l-k}\c_3^k, \nn
C_{3,n} &=&
 - \c_- \left[ \c_0^{n-1}
 + \sum_{l,k} \frac{l}{(l!)^2} \frac{(k+l-1)!(n-1-k-l)!}{k!(n-1-k-2l)!}
 (\c_+ \c_-)^l \c_0^{n-1-2l-k}\c_3^k \right] C_{m,+}, \nn
&& + \c_+ \left[ \c_3^{n-1}
 + \sum_{l,k} \frac{l}{(l!)^2} \frac{(k+l-1)!(n-1-k-l)!}{k!(n-1-k-2l)!}
 (\c_+ \c_-)^l \c_3^{n-1-2l-k}\c_0^k \right] C_{m,-}, \nn
&& + \sum_{l,k} \frac{2}{(l!)^2} \frac{(k+l)!(n-2-k-l)!}{k!(n-2-k-2l)!}
 (\c_+ \c_-)^{l+1} \c_0^{n-2-2l-k}\c_3^k  C_{m,3} \nn
&& -2 (\pa \c_0) \c_0^{n-1} + 2 (\pa\c_3)\c_3^{n-1}
+ \sum_{l,k} \frac{2 l}{(l!)^2} \frac{(k+l-1)!(n-1-k-l)!}{k!(n-1-k-2l)!}
 (\c_+ \c_-)^l \nn
&& \qquad \qquad \times \left[ (\pa\c_3)\c_3^{n-1-2l-k}\c_0^k -
 (\pa\c_0)\c_0^{n-1-2l-k}\c_3^k \right] \nn
&& + \sum_{l,k} \frac{2}{(l!)^2} \frac{(k+l)!(n-2-k-l)!}{k!(n-2-k-2l)!}
 (\c_+ \c_-)^l \left[ (\pa\c_+)\c_-\c_3^{n-2-2l-k}\c_0^k \right. \nn
&& \left. \qquad \qquad - \c_+(\pa\c_-)\c_0^{n-2-2l-k}\c_3^k \right],
\ena
where the sums over $k$ and $l$ run over non-negative integers.

The large $N=4$ BRST operator is given by
\bea
Q_{N4} &=& \int \frac{dz d\t}{2\pi i} \left[
c_t T + \c_{c+} C_+ + \c_{c-} C_- + \c_{c3} C_3 + c_{n+} N_+
 + c_{n-} N_- + c_{n3} N_3 + c_{h} H  \right. \nn
&& + c_t \left[ \shalf T_{3/2}(\b^t,c_t) + T_1(b^{c+},\c_{c+})
 + T_1(b^{c-},\c_{c-}) + T_1(b^{c3},\c_{c3}) + T_{1/2}(\b^{n+},c_{n+})
 \right. \nn
&& \left. + T_{1/2}(\b^{n-},c_{n-}) + T_{1/2}(\b^{n3},c_{n3}) +
 T_{1/2}(\b^{h},c_{h}) \right] \pm 2(D\c_{c3})b^{c\pm}\c_{c\pm}
 \pm \c_{c3}(Db^{c\pm})\c_{c\pm} \nn
&& - \b^t(\c_{c3})^2 + 4(D\c_{c+})b_{c3}\c_{c-} - 4 \c_{c+}\b^t \c_{c-}
 +2\c_{c+}(Db^{c3})\c_{c-} +(Dc_{n3})\b^h \c_{c3} \nn
&& \pm (Dc_{n3})\b^{n\pm}\c_{c\pm} \pm c_{n3}b^{c\pm}\c_{c\pm}
 \pm (Dc_{n\pm})\b^{n\pm}\c_{c3} \pm c_{n\pm}b^{c\pm}\c_{c3}
 - 2(Dc_{n\pm})(\b^h \mp \b^{n3})\c_{c\mp} \nn
&& \left.  \mp 2 c_{n\pm}b^{c3}\c_{c\mp} \mp c_{n3}\b^{n\pm}c_{n\pm}
 -2 c_{n+}\b^{n3}c_{n-} - Dc_h \b^{n\pm}\c_{c\pm}
 + Dc_h \b^{n3}\c_{c3} \right],
\label{n4brs}
\ena
where the ghosts for each generators are labelled by their corresponding
subscripts and superscripts, and
\EQ
T_j(b,c) = (-1)^{\e_b+1}j b\pa c + (-1)^{\e_b}\left(\shalf-j \right)
\pa b c + \shalf Db Dc,
\label{emten}
\EN
is the energy-momentum tensor for the $(b,c)$-system of spin
$(j,\shalf-j)$. In eq.~(\ref{emten}), the statistics of the fields
$(b,c)$ are not restricted except that $b$ and $c$ have opposite
characters, and $\e_b$ is odd (even) for anticommuting (commuting)
$b$. Substituting the generators (\ref{n1gen}) into (\ref{n4brs}),
we obtain the BRST operator for this system which contains an
infinite number of terms.

We now show that the BRST operator can be mapped by a series of
similarity transformations into the sum of those
for the small $N=4$ string $Q_{sN4}$ and topological sector $Q_{top}$:
\EQ
Q_{sN4} + Q_{top},
\EN
where
\bea
Q_{sN4} &=& \int \frac{dz d\t}{2\pi i} \left[
c_t T + \c_{c+} C_+ + \c_{c-} C_- + \c_{c3} C_3 \right. \nn
&& + c_t \left( \shalf T_{3/2}(\b_t,c_t) + T_1(b^{c+},\c_{c+})
 + T_1(b^{c-},\c_{c-}) + T_1(b^{c3},\c_{c3}) \right) \nn
&& \pm 2(D\c_{c3})b^{c\pm}\c_{c\pm} \pm \c_{c3}Db^{c\pm}\c_{c\pm}
 - \b^t(\c_{c3})^2 + 4(D\c_{c+})b^{c3}\c_{c-} \nn
&& -\left. 4\c_{c+}\b^t \c_{c-} + 2 \c_{c+}(Db^{c3})\c_{c-} \right], \nn
Q_{top} &=& \int \frac{dz d\t}{2\pi i} \left[ c_{n+} b^{-}
 + c_{n-} b^{+} + \shalf c_{n3}(b^3 - b^0) - \shalf c_{h}(b^3 + b^0)
\right].
\ena

To show this, we first transform it by
\bea
R_1 &=& \int\frac{dzd\t}{2\pi i} \left[
\{ \c_{c+}(\c_3-\c_0) + \c_{c3}\c_- \} b^{c+}
- \{ \c_{c-}(\c_3-\c_0) + \c_{c3}\c_+ \} b^{c-} \right. \nn
&& + 2 (\c_{c+}\c_+ - \c_{c-}\c_-) b^{c3}
- (\c_3 + \c_0)(\c_+ b^+ + \c_- b^-)
- (\c_0^2 + \c_+ \c_-) b^0 - (\c_3^2 + \c_+ \c_-) b^3 \nn
&& + \left\{ \c_0 c_{n+} +\shalf \c_- (c_{n3} - c_h) \right\} \b^{n+}
 + \left\{ \c_3 c_{n-} - \shalf \c_+ (c_{n3} + c_h) \right\} \b^{n-} \nn
&& + \left\{ \c_+ c_{n+} - \c_- c_{n-} + \shalf \c_3 (c_{n3} - c_h)
 + \shalf \c_0 (c_{n3} + c_h) \right\} \b^{n3} \nn
&& \left. - \left\{ \c_+ c_{n+} + \c_- c_{n-} + \shalf \c_3 (c_{n3}
 - c_h) - \shalf \c_0 (c_{n3} + c_h) \right\} \b^h \right].
\ena
Remarkably this transformation brings the total BRST operator into
a sum of finite number of terms. The next transformation generated by
\bea
R_2 &=& \int\frac{dzd\t}{2\pi i} \left[
\{ \c_{c3}(\b^{n3}+\b^h) - 2 \c_{c-}\b^{n-} \} D\c_0
+ \{ \c_{c3}(\b^{n3}-\b^h) - 2 \c_{c+}\b^{n+} \} D\c_3 \right. \nn
&& \left. + \{ \c_{c3}\b^{n-} + 2 \c_{c+}(\b^{n3}+\b^h) \} D\c_+
- \{ \c_{c3}\b^{n+} + 2 \c_{c-}(\b^{n3}-\b^h) \} D\c_- \right],
\ena
then eliminates irrelevant cubic terms from the BRST operator.
A final transformation by
\bea
R_3 &=& \int\frac{dzd\t}{2\pi i} \shalf c_t [ \b^{n+}\pa\c_-
 - D\b^{n+}D\c_- + \b^{n-}\pa\c_+ - D\b^{n-}D\c_+
 + \b^{n3}\pa(\c_3-\c_0) \nn
&& - D\b^{n3}D(\c_3-\c_0) - \b^{h}\pa(\c_3+\c_0) + D\b^{h}D(\c_3+\c_0)],
\ena
maps it to the sum of $Q_{sN4} + Q_{top}$. In the same manner as the
$N=2$ string, the BRST operator $Q_{top}$ imposes the condition that
the fields $(c_{n\pm},\b^{n\pm},\c_{\mp},b^{\mp}),
(c_{n3},\b^{n3},\c_3-\c_0,\shalf(b^3-b^0)),
(c_{h},\b^{h},\c_3+\c_0,\shalf(b^3+b^0))$ all fall into the quartet
representation of the BRST operator and hence decouple from the
physical sector. Thus the large $N=4$ superstring in our particular
background is equivalent to the small $N=4$ superstring.

Finally let us discuss if the general large $N=4$ superstrings can be
embedded into higher superstrings in the spirit of the universal string.
It has been shown in ref.~\cite{BO1} that the critical condition for
the large $N=4$ superstring is that the total central charge is zero
but that the ratio $x$ of the levels of the two $SU(2)$ current
algebras in its SCA can be arbitrary. In ref.~\cite{BOP}, those with
the equal levels were embedded into the hierarchy of superstrings.
The question thus remains if the general large $N=4$ superstring with
$c=0$ but an arbitrary $x$ can be embedded into the $N=5$ superstrings.
However, as can be seen from eq.~(\ref{n1large}), the critical $N=4$
SCA can be written in the same form for an arbitrary ratio of the levels
just by changing the correspondence with the component generators. We
have checked that this is true also when it is written in $N=2$
superfields. Hence after this appropriate redefinitions of the
superfield generators, the realization given in ref.~\cite{BOP} is
actually valid for an arbitrary ratio of the levels.

This completes the program of embedding all the
superstrings into the unified framework of the universal string.

\vs{5}
\noindent
{\it Acknowledgements}

We have checked many of our super-OPEs using the Mathematica package
developed by S. Krivonos and K. Thielemans, whose software is
gratefully acknowledged.
We would also like to thank H. Kunitomo for valuable discussions.
In addition, we thank J. O. Madsen and J. de Boer for useful advice
on the use of OPE package.

\end{document}